\documentclass[twocolumn,letterpaper]{revtex4-1}
\usepackage[english]{babel}
\usepackage{graphicx}
\usepackage{epsfig}
\usepackage{epstopdf}
\usepackage{sidecap}
\usepackage[normalem]{ulem}
\begin{document}

\begin{abstract}
Transmitting messages in the most efficient way as possible has always been one of politicians main concerns
during electoral processes. Due to the rapidly growing number of users,
online social networks have become ideal platforms for politicians to interact with their
potential voters. Exploiting the available potential of these tools to maximize their influence over voters
is one of politicians actual challenges. To step in this direction, we have analyzed the user activity in the online social network
Twitter, during the 2011 Spanish Presidential electoral process, and found that such activity is correlated with the election results.
We introduce a new measure to study political support in Twitter, which we call the Relative Support. 
We have also characterized user behavior by analyzing the structural and dynamical patterns of the complex networks emergent
from the mention and retweet networks. Our results suggest that the collective attention is driven by a very small fraction
of users. Furthermore we have analyzed the interactions taking place among politicians, observing a lack of debate. Finally
we develop a network growth model to reproduce the interactions taking place among politicians.
\end{abstract}

\title{Characterizing and modeling an electoral campaign in the context of Twitter: 2011 Spanish Presidential Election as a case study}


\author{J. Borondo}
 \affiliation{Grupo de Sistemas Complejos and Departamento de F\'isica y Mec\'anica. Universidad Polit\'ecnica de Madrid. ETSI Agr\'onomos, 28040, Madrid, Spain}
 \author{A. J. Morales}
  \affiliation{Grupo de Sistemas Complejos and Departamento de F\'isica y Mec\'anica. Universidad Polit\'ecnica de Madrid. ETSI Agr\'onomos, 28040, Madrid, Spain}
  \author{J. C. Losada}
 \affiliation{Grupo de Sistemas Complejos and Departamento de F\'isica y Mec\'anica. Universidad Polit\'ecnica de Madrid. ETSI Agr\'onomos, 28040, Madrid, Spain}
 
 \author{R. M. Benito}
 \affiliation{Grupo de Sistemas Complejos and Departamento de F\'isica y Mec\'anica. Universidad Polit\'ecnica de Madrid. ETSI Agr\'onomos, 28040, Madrid, Spain}
\maketitle



\section{Introduction}
The electoral campaign is a period preceding elections where political parties do an organized 
effort so that their candidates garner supporters. Maximizing the influence of their messages over voters is the
main objective. In this way, politicians use different techniques to transmit their messages in the
most effective way to their potential voters, such as mass meetings, rallies, husting or media 
management. Understanding and exploiting in a more efficient way the available resources for 
information flow than your opponent can make the difference.

 Over the last century mass media has been monopolized by 
\lq\lq{}old media\rq\rq{}, such as televisions or newspapers. However nowadays we are attending to a
transition where a new interactive online social media world is settling its bases. Online social networks, such as 
Twitter with over 200 million users,
have become ideal platforms for information flows. This has been noted in \cite{Courtenay:f-I5M7De} 
where they reported that these tools may serve as a framework for discussion. Other studies
have been directed 
towards identifying influential users \cite{Romero:2010tj} or discovering its commercial 
usage \cite{Janse:gu}.
Moreover the percentage of population using online social networks has
increased in recent years, reaching in Spain a $42\%$ of the population, quantity that is almost
duplicated ($82\%$) for young adults between 18-29 years old \cite{pew}.

Following the idea \lq\lq{}one must be where people are\rq\rq{}, politicians are now present in the most 
popular online social networks. However some politicians do not have a defined
strategy for the usage of these tools and the rest are still far of exploiting all the available potential. 
The importance and popularity of social media in politics became clear with Obama's campaign for
 the
2008 U.S. Presidential elections and his famous tweet: "This is history...", posted just after winning 
the elections. This fact attracted not only popular, but also scientific attention, making political 
conversations in Twitter a popular subject for research. Lately, the data gathered from
Twitter has been used as a \lq\lq{}social 
sensor\rq\rq{} to predict election outcomes \cite{alemanas}. Other studies have focused in analyzing the 
interactions between different political communities \cite{pol_polar}, and finally a 
proof-of-concept-model has been developed \cite{adamic_model} to predict candidate's victory.

In this article we introduce a new parameter that measures the ratio of the support in Twitter between
two candidates, which we call the Relative Support, and apply it to the 2011
Spanish Presidential elections, to show how it can be used to indicate and quantify which candidate
and in which proportion is getting more benefits from events occurring offline. We further study the
dynamical patterns emergent from the Twitter mention and retweet 
networks within the framework of complex networks theory \cite{BocalettiReview, BarabasiReview, NewmanReview}. We also interpret politicians behavior by filtering these networks and analyzing
the interactions going on between the different political parties. Finally we introduce a model based
on the heterogeneous preferential attachment formalism \cite{Santiago2007} capable of growing political conversations and illustrate it by reproducing the mentions and retweets taking place in Twitter among politicians.

\section{System}

The present work is based on data collected from the online social network Twitter. This web application allows people to post and exchange text messages limited by $140$ characters. 
There are several interaction mechanisms in Twitter to transfer information. The first of these is the ability of people to follow and be followed by
other users. This is a passive mechanism that allows users to receive the messages written by
their followees in real time. The Twitter's followers network is a directed graph where non reciprocal
relations are admitted and it states the social substratum through which information may flow. Previous studies have reported a high heterogeneity in the followers distribution \cite{Kwak}. Another important mechanism to interact is the retweet or message retransmission. This
mechanism allows individual messages to propagate throughout the network and serves as a way for people to endorse their point of view over specific subjects
\cite{DanahTweetRetweet}. In addition to this, another relevant way for direct interaction is the mentions mechanism. By mentioning someone's username in the message text, people is able to send directed messages to the mentioned user's inbox. This mechanism is often used to establish conversations between users, through the exchange of messages, or just to refer somebody in the message's text \cite{Courtenay:f-I5M7De}. 

Our dataset is constructed from public access messages posted in Twitter, related to the 2011
Spanish Presidential elections. We downloaded all the messages that included the keyword
{\it 20N}, using the Twitter API interface, in a three week period including the election day. We chose
this keyword because is an ideologically neutral identifier, used by all the political parties during the campaign and voting day. In
summary we analyzed over 370.000 messages, written by over 100.000 users. We found that $40\%$ of the
messages were retweets, and over $25\%$ contained at least one mention. This fact makes the
event quite relevant, since it has been reported that retweets represent about $4\%$ of the overall messages \cite{pearnalitycs}.\\
\section{Results and Discussion}

\subsection{Time Series}

\begin{table*}
\caption{Results by political party for the votes obtained, the mentions on tweets and the messages sent from official accounts (Activity).}
\label{Votes}
\begin{tabular}{ |c|c|c|c|c|}
\hline \textbf{Political Party}& \textbf{Acronym} & \textbf{$\%$ Votes} & \textbf{$\%$ Tweets} & \textbf{Activity}\\
\hline Partido Popular & PP & 44,62 & 39,92 & 1228 \\
\hline Partido Socialista Obrero Espa\~nol & PSOE & 28,73 & 26,33 & 1819\\
\hline Izquierda Unida &IU & 6,92 & 5,03 & 451 \\
\hline Uni\'on Progreso y Democracia & UPyD & 4,69 & 11,8 & 1852 \\
\hline Convergencia i Unio & CIU & 4,17 & 4,51 & 208 \\
\hline AMAIUR & AMAIUR & 1,37 & 2,76 &  11 \\
\hline Partido Nacionalista Vasco & PNV & 1,33 & 2,20 & 11\\
\hline Ezquerra Republicana de Catalunya & ERC & 1,05 & 1,47 & 113\\
\hline
\end{tabular}
\end{table*}
We can begin to understand how the Spanish political landscape is reflected in Twitter by 
comparing the number of times that each political party has been mentioned during the {\it 20N}
discussion and the number of votes it obtained in the elections. Previous studies 
show that there is a correlation between the number of times a political party is mentioned 
during an electoral campaign on  Twitter and the number of votes the political party obtains \cite{alemanas}. 
These results are backed up by our study, where we find tweets to be a quite accurate survey. We prove
this statement by
ordering the political parties with at least a $1\%$ of votes by the number of 
votes they obtained and comparing it to the number of times they were mentioned during the {\it 20N} conversation. 
The results are presented in Table \ref{Votes}, where the name of these parties and their acronyms can be found. We observe that the only deviation from the predicted 
order is the swap of positions between UPyD and IU. Despite IU obtaining more votes,
UPyD was mentioned more times. This can be explained by the 
much more active Twitter campaign done by UPyD in comparison to IU, that barely used this media to campaign, as it can be seen in Table \ref{Votes}.

\begin{center}
\begin{figure*}
\includegraphics[width=500px]{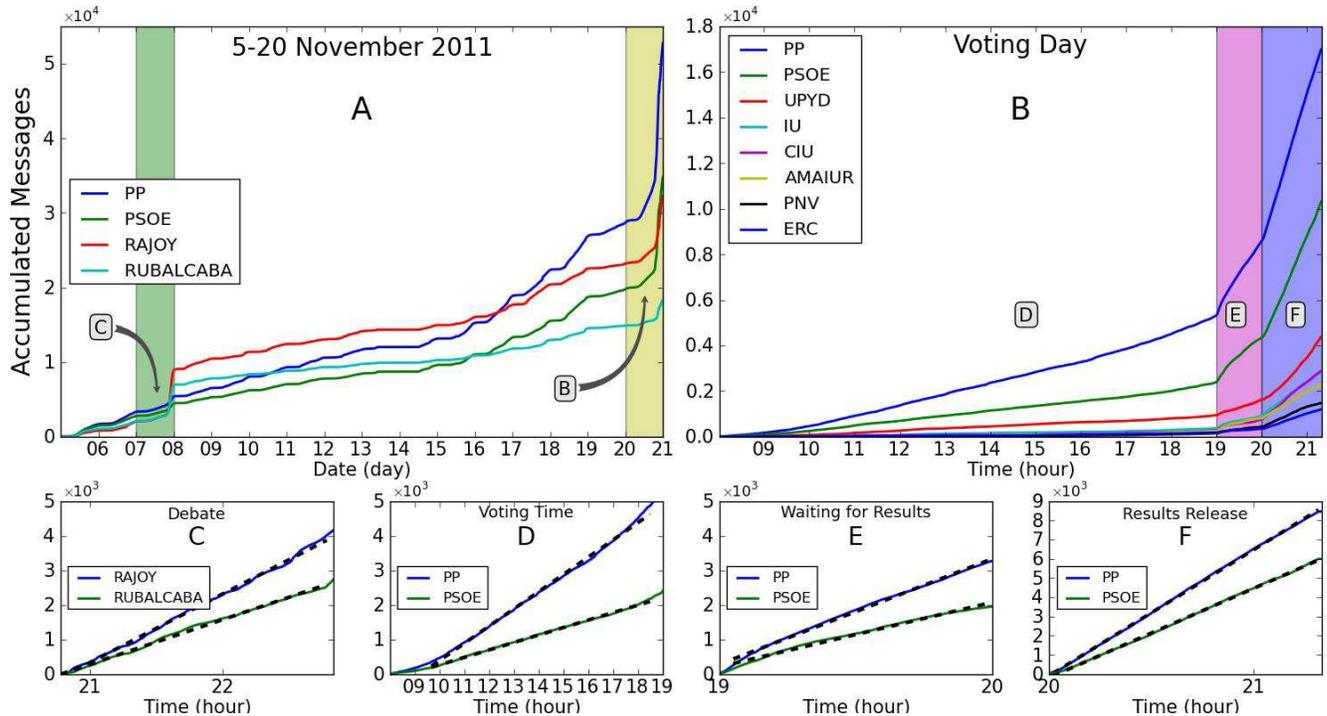}
\caption{Time series of the accumulated tweets mentioning political parties and candidates for the entire campaign (A), voting day (B), debate (C), voting time (D), waiting for results (E) and results release (F). The dashed lines (C, D, E, F) represent a linear fit. At all panels, the order of the labels in the legend corresponds to the same order of the curves at their final value in descendant way. In the horizontal axis, hours are given in UTC time.}
\label{Series}
\end{figure*}
\end{center}

Since in Spain there are two main political parties that outstand on top of the others, we
focused our study on them. We analyze the time series of the accumulated tweets mentioning 
at least one of these parties, PP and PSOE, or their candidates, Rajoy 
and Rubalcaba. Looking at Figure \ref{Series}A we can state two things: Firstly tweets contain more mentions to the political parties rather than to their candidates; secondly the more conservative 
party, PP and its candidate Rajoy, were much more mentioned than PSOE and Rubalcaba.

One of the most important results we have obtained studying the {\it 20N}  Twitter 
conversation, is that the time series of the accumulated tweets mentioning political parties or 
candidates
present piecewise linear growth, as it is showed in Figure \ref{Series}B. On top of that the points where the slope increases coincides
with important events occurring outside Twitter that fuel the activity of the conversation, which occur at the same time for both parties.
This fact makes us think about the ratio between the rate at which the cumulative mentions to two
political parties grow as a better indicator of the outside political support to each party, than just the raw
number of mentions. Following this idea we define an instant indicator of the support in Twitter between two political parties, which we call the Relative Support parameter
$RS_{B}^{A}$, given by the following expression:
\begin{equation}\label{Politicial Support}
RS_{B}^{A}=\frac{m_{A}}{m_{B}}
\end{equation}
where $m_{A}$ and $m_{B}$ are the slopes for the accumulated mentions to the $A$ and $B$ political parties.

From our point of view there are two days of special relevance in our study: the debate between the two main candidates that took place on November 7th, and the 
voting day on November 20th. We did a further analysis of these two days.

During the debate, people's attention was completely focused on the two candidates, Rajoy and
Rubalcaba. This provoked that, contrary to what happened during the whole campaign, the 
candidates were more mentioned than their corresponding political parties, as it can be seen in
Figure \ref{Series}A. Therefore, for this period, we studied the time series of the accumulated mentions to the candidates
rather than the parties. The majority of tweets
about {\it 20N} posted on this day are concentrated on the two hours that lasted the debate, with a total of 2.733 messages mentioning Rubalcaba and 4.150 mentioning Rajoy. In Figure
\ref{Series}C we present a detail of the time series of the accumulated tweets for the candidates 
during the debate. We can observe that, in accordance with what we said before, both series present linear growth, 
being the slopes for both candidates constant during the whole debate, and changing its value at 
the end of it. The Relative Support  during the debate was
$RS^{Raj}_{Rub}=1,53$, Rajoy over Rubalcaba. This value is pretty close to the relation 
between the votes (1,55) obtained by the two candidates thirteen days after (see Table \ref{RPS}).

\begin{table*}
\caption{Comparison between the ratio of votes and the Relative Support parameter for the two main political parties.}
\label{RPS}
\begin{tabular}{ |c|c|c|c|c|}
\hline \textbf{ Votes Ratio} &  \textbf{Debate} & \textbf{ Voting Time} &  \textbf{Waiting for Results} & \textbf{ Results Release}\\
\hline $\frac{PP}{PSOE}=1,55$ & $RS^{PP}_{PSOE}=1,53$ & $RS^{PP}_{PSOE}=2,31$ & $RS^{PP}_{PSOE}=1,64$ & $RS^{PP}_{PSOE}=1,41$\\
\hline
\end{tabular}
\end{table*}

The election day survey is one of the most relevant and reliable surveys to predict election outcomes.
This makes us believe
that a further study on this day must be done when analyzing election results. As it can be seen in
Figure \ref{Series}A, the major increase of political mentions occurred during the voting day, what reinforces
our idea about the importance of this day. In Figure \ref{Series}B we show
a detail of the time series of the accumulated messages from 8:00 to 21:20 for the Spanish political 
parties. In correspondence with our theory of piecewise linear
growth, in Figure \ref{Series}B we can distinguish three important regions (D, E, F) of the space-time for this day: 
\lq\lq{}Voting time\rq\rq{}, \lq\lq{}Waiting for results\rq\rq{}, \lq\lq{}Results release\rq\rq{}, that we further discuss, and present in panels D, E and F.

VOTING TIME (8:00-19:00). This panel covers the entire voting period, from the opening of the electoral 
colleges to the closure. Over 7.500 tweets containing either PP or PSOE were posted. From the four panels studied in detail (Figure \ref{Series}C, D, E, F), this one presents by far the lowest activity per hour, what makes it the less representative sample. The Relative Support 
took a value of $RS^{PP}_{PSOE}=2,31$ in favor of PP, indicating that PP users were much more 
enthusiastic than PSOE.

WAITING FOR RESULTS (19:00-20:00). This period lasts only one hour, starting with the closure of polls 
and ending when the first news were released. This first news informed about the 
participation statistics, and gave provisional results for a $5\%$ scrutiny. Over 5.000 tweets mentioning either of the two main parties were posted during this hour. During this period the Relative Support  parameter estimated quite accurately the upcoming results, taking a value of, $RS^{PP}_{PSOE}=1,64$ in favor of PP. 

RESULTS RELEASE (20:00-21:30). This region covers the entire period in which the results were
given, starting with a $5\%$ of scrutiny and ending with an $85\%$, point at which the politicians 
made their first speeches. It was the period with more activity per hour in Twitter of the whole 
study, with more than 13.000 tweets posted mentioning PP or PSOE. The measure of the Relative Support 
while results were given was of $RS^{PP}_{PSOE}=1,41$, pretty close to the relation between the 
votes of the two parties, as it can be seen in Table \ref{RPS}.

Summarizing, we have centered our study on the two dominant parties of the Spanish political
landscape
and observed that in this system the relation in votes and tweets between them coincides quite 
precisely (Table \ref{Votes}). We introduce a new measure to study political support in Twitter, the 
Relative Support between two parties $RS_B^A$, which we see as a useful
tool to study 
future elections or to determine how Twitter users react to external events, and who gets more popular with them.
In our study we identify the debate between the two candidates (7th of November) as the key point 
in Twitter. This was the point where users of the social network began to actively participate in the
{\it 20N} conversation, and during the two hours of debate people reflected their preferences in Twitter,
$RS_{PSOE}^{PP} = 1,54$. The lack of external critical political events during the campaign and
the firmness of people's vote intention, maintained the ratio of tweets constant around this
value along the whole campaign. 
Although future work should be done in applying the RS parameter to other elections, we believe that this parameter is
capable of revealing election outcomes even when offline events occurring at the last minute change
voter support. In this way it would have detected Zapatero's victory against forecast in the 2004 Spanish
Presidential election, that took place four days after the 11M terrorist attack.

\subsection{User Interactions}
\begin{center}
\begin{figure*}
\includegraphics[width=500px]{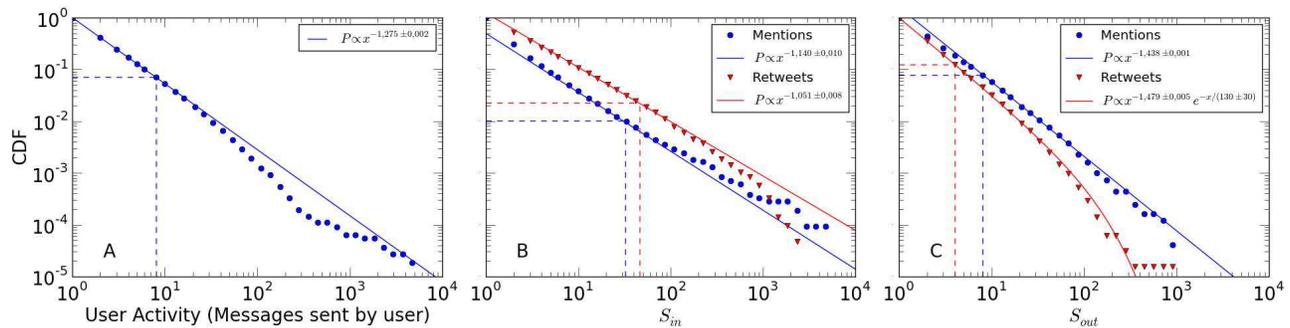}
\caption{User cumulative distribution for user activity (A), mention and retweet networks in strength (B) and mention and retweet networks out strength (C). The solid lines represent the best fitted curve for each distribution. The dashed lines indicate the percentage of users that posted $50\%$ of the messages (A), received $50\%$ of the mentions or retweets (B) and made $50\%$ of the mentions or retweets (C).}
\label{Dinamica}
\end{figure*}
\end{center}

So far we have seen that the user activity is correlated with the election results. In this section we will further analyze such activity and characterize its emergent structural and dynamical patterns based on the Twitter interaction mechanisms. First of all, we have analyzed the cumulative probability distribution for the user activity, that we define as the number of messages posted by user. This distribution follows a power law in the form $P(x>x^*) = x^{- \beta}$, where $\beta = 1,275 \pm 0,002$, as shown in Figure \ref{Dinamica}A. Such exponent is within the expected values for scale-free human activity phenomena \cite{newman2005power}. This implies a very high heterogeneity level in user behavior. In fact, we found that half of the messages were posted by only $7\%$ of the participants, who were the most active users and posted from 8 to over 4.000 messages each, while the other half of the messages were posted by the remaining $93\%$ of users, who posted less than $8$ messages each. Similar results were obtained in the study of the 2009 German elections \cite{alemanas} and in the study of the 2005 Canadian elections \cite{canadienses}, who concluded that the political discussions during the campaign in social media were controlled by a very small fraction of the participants. However, it is unclear whether this activity really represented an actual discussion or debate. To answer this question we will next analyze the user activity taking into account the way participants interacted with each other, either by the mention or retweet mechanisms. Therefore we have built two networks according to \lq\lq{}who mentioned who\rq\rq{} and \lq\lq{}who retweeted (or retransmitted) who\rq\rq{}. Both networks have directed and weighted edges \cite{barrat2004weighted, BocalettiReview}, whose weight is directly proportional to the number of times that a user has mentioned or retweeted another user. In total, the mention network has over 39.631 nodes and 86.029 links, while the retweet network has over 75.546 nodes and 153.549 links. In Table \ref{NetProperty} we present the networks' main properties, some of which we will discuss next

In Figure \ref{Dinamica}B we present the in strength cumulative distribution for both networks. The in strength indicates the number of mentions received by user, and the number of retweets gained by user, respectively. Both measures are related to the level of collective attention that users may gather along the conversation. The in strength distributions follow power laws in the form $P(x>x^*) = x^{-\beta}$ where $\beta_{M}=1,14 \pm 0,01$ and $\beta_{R}=1,051 \pm 0,008$. Once more, such distributions display a high heterogeneity level found in the users profiles. As a matter of fact, we found that just $1,04\%$ of the users were target for half of the total mentions and $2,24\%$ of the users wrote the messages that caused half of the total retransmissions. These results show that both mechanisms are highly elitist, since a remarkably small fraction of users, mainly compound by media and politicians, concentrate half of the collective attention, while the large majority individually attracted only a few. Such collective attention is built out of adding individual efforts, which are characterized in the out strength distributions, shown in Figure \ref{Dinamica}C. These distributions indicate the amount of mentions or retransmissions made by user, respectively. For the mention network we found a power law distribution in the form $P(x>x^*) = x^{- \beta}$ where $\beta_{M}=1,438 \pm 0,001$, and for the retweet network we found that the data fit better to an exponentially truncated power law in the form $P(x>x^*) = x^{- \beta}e^{-x/c}$ where $\beta_{R}=1,479 \pm 0,005$ and $c = 130 \pm 30$. As we found on the overall user activity distribution, the out strength distributions show that a small fraction of users ($7,71\%$ for mentions and $12,41\%$ for retweets) concentrated over half of the activity, while the majority of users who concentrated the other half ($92,29\%$ for mentions and $87,59\%$ for retweets), mentioned less than 8 users and retweeted less than 4 messages.

\begin{table}
\caption{Topological properties of the mention and retweet networks. $r$ represents the assortativity by degree coefficient combined by in and out degrees.}
\begin{tabular}[c]{|c|c|c|}
\hline \textbf{Property} & \textbf{Mentions} & \textbf{Retweets}\\
\hline Nodes & $39.631$& $75.546$\\
\hline Edges & $86.029$ & $153.549$\\
\hline Bidirectional Edges& $2,17\%$ & $0,99\%$\\
\hline $r_{out,out}$& $ -0,039$ & $0,087$\\
\hline $r_{out,in}$& $-0,141$& $-0,107$\\
\hline $r_{in,in}$& $-0,021$ & $-0,043$\\
\hline $r_{in,out}$& $-0,005$& $0,017$\\
\hline
\end{tabular}
\label{NetProperty}
\end{table}

In order to unveil how such heterogeneous users interacted with each other, we have also calculated the assortativity by degree coefficient for both networks \cite{Newman2003a}. As our networks have directed edges, we have calculated this measure by splitting it into combinations of in and out degree pairs \cite{Foster2010}. As shown in Table \ref{NetProperty}, we found the out-in and the in-in pair to be slightly disassortative for both networks ($r_{out,in}^M = -0,141$, $r_{out,in}^R =-0,107$, $r_{in,in}^M = -0,021$, $r_{in,in}^R = -0,043$). These results reinforce the asymmetric shape detected of these networks, where the hubs that concentrate much of the incoming links, are often targeted by regular users, who neither mention nor retweet too much, and receive few of the collective attention. The out-out pair seems to be slightly assortative for the retransmission network ($r_{out,out}^R = 0,087$), which implies that users who retransmit a lot, also target users who also retransmit a lot. This result is related to the fact that retransmissions occur in cascades \cite{outtweeting}, contrary to mentions that do not imply an explicit propagation process, and actually presents a minor degree of disassortativity ($r_{out,out}^M = -0,039$). Finally, we found the in-out pair to be a little assortative for the retransmission network ($r_{in,out}^R = 0,017$), and practically not correlated for the mention network ($r_{in,out}^M = -0,005$). This implies that the highly targeted people, do not tend to target an specific kind of user.

Previous works on network assortativity \cite{Newman2003a}, state that social networks tend to be assortative, as popular people want to be friend with popular people, and regular people are usually friends among the regular people. However our measures indicate the opposite. This was already reported by Hu and Wang \cite{Hu2009osn_official}, who detected that most online social networks are disassortative and in the same order as the networks of our study. The reason for this result is that online relations are different from real life ones. For example in Twitter, regular people are now able to relate and communicate with popular accounts, either by following, mentioning or retweeting their messages. This new kind of interactions are responsible for the changes in the structural and dynamical patterns previously reported on social networks.

\begin{center}
\begin{figure}
\includegraphics[width=250px]{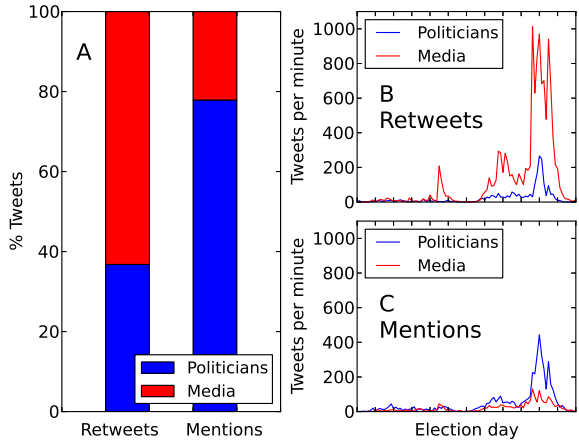}
\caption{Comparison of the percentage of retweets and mentions targeted to politicians and mass media official accounts (A). Comparison of the rate of retweets (B) and mentions (C) targeted to politicians and mass media official accounts during the voting day.}
\label{rm}
\end{figure}
\end{center}

We have also carried out a community structure analysis for both networks based on a random walk algorithm \cite{MapEquation} and found that this conversation also presents a modular structure. In fact the largest modules in the mention network are formed around politicians, while mass media accounts centered the largest modules in the retweet network. In an effort to further discover the role that politicians and mass media accounts have played during the election campaign, we have analyzed how users mentioned and retweeted these accounts. We found that most mentions were targeted to politicians ($77,83\%$) while the most retweeted were mass media accounts ($63,24\%$), as it can be seen in Figure \ref{rm}A. The importance of mass media accounts in the retweet network becomes even more clear during important offline events that produce activity bursts (e.g. Election Day shown in the panel B of Figure \ref{rm}). Therefore we confirm that retweets are used by users to propagate news, originally posted by the traditional media, and to endorse individual opinions.

In summary, we found that most of people participated by posting only a couple of messages that were mostly targeted to a minority part of the participants.
Such elitist group is mainly compound by popular accounts, like media and politicians, and provokes the emergence of communities around them due to their high influence.
Furthermore, most of the interactions occurred in one way only, since just a small fraction of the edges, $2,17\%$ for mentions and $0,99\%$ for retweets, were bidirectional (see Table \ref{NetProperty}). This lead us to suggest that users do not actually discuss with each other. In fact, the interaction mechanisms were mainly used to campaign rather than debate, as we will see in the following section.

\subsection{Political Interactions}
\begin{center}
\begin{figure}
\includegraphics[width=250px]{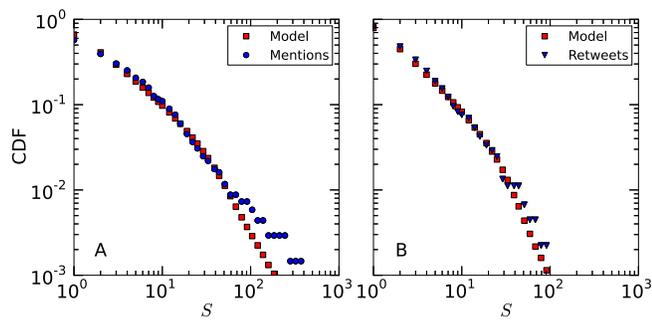}
\caption{Comparison of the strength function cumulative distribution for the filtered by politicians and political parties official accounts mention (A) and retweet (B) networks (Blue) and the results obtained for the proposed model (Red).}
\label{Modelo}
\end{figure}
\end{center}

In order to more deeply understand the political landscape in Twitter,
we have analyzed the political filtered mention and retweet networks. Our target is to find patterns
that help us understand how politicians interacted during the campaign.
For this matter we have filtered the previously mentioned networks, only remaining the official
politician's accounts from those parties with over 10 participants in the conversation. We found that politicians do not relate to each other randomly and that they
use the different interaction mechanisms with specific purposes. The cumulative strength function distribution of both networks are shown in Figure \ref{Modelo}.

In order to understand the way politicians used both mechanisms, we have analyzed the mention and retweet networks by measuring assortative mixing patterns according to discrete characteristics. Therefore we have classified all nodes according to the political party they belong to, and calculated the assortativity coefficient $r$ over the matrix $e_{ij}$, that defines the fraction of edges going from one party to another, as described by Newman in \cite{Newman2003a}. We found $r$ to be close to 1 at both networks ($r_M = 0,905$ and $r_R = 0,990$), as presented in Table \ref{NetPropPolit}, which means
that politicians mentioned and retweeted mostly their own partisans. However, mentions tend to
happen across parties a little more than retweeting, which is the most segregative interaction
found. This result indicates a considerable lack of debate between the politicians and reveals some
of the strategies followed by them during the campaign.
A previous work on Korean elections \cite{koreanas}, reported that mentions between politicians
reflect the political alliances between candidates. However we find retweets to be
a more overwhelming interaction to map the political endorsements, as it presents the highest
assortative mixing coefficient. This issue has already been pointed out during the 2010 U.S. Congress elections, where retweets were found to be more ideologically polarizing than mentions among regular users \cite{pol_polar}.

\begin{table}
\caption{Comparison of the assortative mixing by political party for the mention and retweet networks filtered by politicians official accounts and the proposed model results.}
\label{NetPropPolit}
\begin{tabular}[c]{|c|c|c|}
\hline \textbf{Network} & \textbf{Experimental r} & \textbf{Modeled r}\\
\hline Mention & $0,905$&  $0,86 \pm 0,03$\\
\hline Retweet & $0,991$ & $0,989 \pm 0,005$\\
\hline
\end{tabular}
\end{table}

To further explain the structural features found in the interactions between politicians, we propose a model based on the heterogeneous preferential attachment formalism \cite{Santiago2007}. The idea behind it, is that the probability of a node $i$ interacting with a node $j$ not only depends on their respective degree, but also of the affinity between them. In our model nodes (politicians) are classified according to discrete characteristics (political parties). Thus the probability of appearance of a new interaction from any politician, $i$ , belonging to party $A$, to a politician $j$,
who belongs to a party $B$, is given by the following expression:
\begin{equation}\label{Number Political Interactions}
P_{ij}=\frac{S_j}{\displaystyle\sum_{j\epsilon B} S_j} f_{AB}
\end{equation}
where $S_j$ is $j$\rq{}s strength, and $f_{AB}$, is the affinity value from $A$ to $B$. The first factor
of equation \ref{Number Political Interactions}  corresponds to the local connection at microscale, and the second one to the mesoscale.
We implement this model in the mesoscale by grouping all the politicians
of the same party in a supernode labeled with the name of the party. The properties of these supernodes are determined by those of the nodes inside them, and the number of interactions, $N$, between them can be obtained from the $e_{ij}$ matrix \cite{Newman2003a}. In this way the affinity value between two politicians will be determined only by their political parties. We represent the directional affinity between political parties by an $f$ matrix whose elements quantify the relative flux of interactions from $A$ to $B$. Using this notation we can model the structure of the $f$ matrix as:
\begin{equation}\label{Affinity Matrix}
f_{AB}= \frac{N_{AB}}{N_{A}}
\end{equation}
where $N_{AB}$ is the total flux going from $A$ to $B$ and $N_{A}$ is the total flux going out from A.
After understanding the structure of the system's mesoscale, it\rq{}s time to model the microscale. In this scale the probability rule for interactions between politicians is based on the preferential attachment model \cite{Barabasi:1999he}. In the sense that the likelihood of
a node, belonging to party $B$, to receive a new interaction, increases with the node\rq{}s strength.
To implement the model we have calculated the experimental $f$ matrix of each network, considering only those parties with over 10 politicians participating in the conversation. Also we have assigned a random number of outgoing edges to the newly added nodes, following power law distributions of exponents: $\gamma_M = 1,3$ for the mention network and $\gamma_R = 1,6$ for the retweets one. In this way we modeled the resulting distributions by simulating the heterogeneity found in the users behavior and using the same microscale connection rule for both interaction mechanisms. In Figure \ref{Modelo} we present the resulting cumulative strength function distribution for both networks, after having averaged over $1.000$ realizations. It can be noticed that the model reproduces perfectly the strength function distribution for both networks, and maintains the assortative mixing levels as presented in Table \ref{NetPropPolit}.
\\
\section{Conclusions}

The perfect political campaign strategy has been eternally chased by politicians. To that effect we have
tracked voter sentiment and uncovered the underlying structure of the campaign in Twitter, measuring 
the impact that different events have produced on politicians popularity and analyzing the roles played
by the various users. For this matter, in this paper we propose a parameter that measures
the relative support in Twitter between two candidates, and apply it to our case of study: the 2011 Spanish 
Presidential elections. Furthermore we have analyzed the graph structural and dynamical 
patterns emergent from interactions taking place among users, finding out that the collective attention is 
driven by a very small fraction of users, who dominate the interaction mechanisms. We have also analyzed politicians behavior finding a profound 
segregation and lack of debate among them. Finally we propose a network growth model based on heterogeneous preferential attachment, 
to explain the emergence of such segregated modules in the politician's networks.
Despite we can't assure that the campaign on Twitter determined the election outcomes,
our results suggest that there is a strong correlation between the activity taking place in Twitter and election results. This fact suggests
that further research should be done on identifying the most efficient techniques to influence voter
sentiment in Twitter.

\begin{acknowledgments}
Support from MICINN- Spain under contracts No. MTM2009- 14621, and i-MATH CSD2006-32, is gratefully acknowledged.
\end{acknowledgments}


\begin{thebibliography}{27}
\bibitem{Courtenay:f-I5M7De} C. Honeycutt and S. C. Herring
{\it HICSS } IEEE Computer Society pp. 1-10 (2009)
\bibitem{Romero:2010tj} D. M. Romero, W. Galuba, S. Asur, and B. A. Huberman 
Influence and passivity in social media.
{\it WWW\rq{}11} (2010)
\bibitem{Janse:gu} B. J. Jansen, M. Zhang, K. Sobel, and A. Chowdury 
{\it Journal of the American Society for Information Science and Technology} 60, 2169 (2009)
\bibitem{pew} Pew Research Center.
Global Digital Communication: Texting, Social Networking Popular Worldwide (2011) 
\bibitem{alemanas} A. Tumasjan, T. O. Sprenger, P. G. Sandner, and I. M. Welpe 
Predicting elections with Twitter: What 140 characters reveal about political sentiment.
{\it ICWSM} The AAAI Press (2010)
\bibitem{pol_polar} M. Conover et al. 
Political polarization on Twitter.
{\it ICWSM} The AAAI Press (2011)
\bibitem{adamic_model} A. Livne, M. P. Simmons, E. Adar, and L. A. Adamic 
The party is over here: Structure and content in the 2010 election.
{\it ICWSM} The AAAI Press (2011)
\bibitem{BocalettiReview} S. Bocaletti, V. Latora, Y. Moreno, M. Chavez and D.-U. Hwang 
{\it Phys. Rep.} 424, 175 (2006)
\bibitem{BarabasiReview} R. Albert and A.-L. Barabasi
{\it Rev. Mod. Phys.} 74, 47-97 (2002)
\bibitem{NewmanReview} M. E. J. Newman 
{\it SIAM Review} 45 2, 167-256 (2003)
\bibitem{Santiago2007} A. Santiago and R. M. Benito 
{\it Europhysics Letters} 82, 58004 (2008)
\bibitem{Kwak} H. Kwak and C. Lee and H. Park and S. Moon 
{\it WWW '10} ACM, 591-600 (2010)
\bibitem{DanahTweetRetweet} D. Boyd, S. Golder and G. Lotan 
{\it HICSS} IEEE Computer Society, pp. 1-10 (2010)
\bibitem{pearnalitycs} pearanalytics. Twitter study (2009)

 http://www.pearanalytics.com/blog/wp-content/uploads/2010/05/Twitter-Study-August-2009.pdf
\bibitem{newman2005power} M. E. J. Newman 
{\it Contemporary Physics} 46, 323 (2005)
\bibitem{canadienses} R. Koop and H. J. Jansen 
{\it Social Science Computer Review} 155 (2009)
\bibitem{barrat2004weighted} A. Barrat, M. Barthlemy, R. Pastor-Satorras, and A. Vespignani 
{\it Proc Natl Acad Sci USA} 101, 3747 (2004)
\bibitem{Newman2003a} M. E. J. Newman 
{\it Physical Review E} 67, 026126 (2003)
\bibitem{Foster2010} J. G. Foster, D. V. Foster, P. Grassberger, and M. Paczuski 
{\it Proc Natl Acad Sci USA} 107, 10815 (2010)
\bibitem{outtweeting}K. A. Wojciech Galuba, D. Chakraborty, Z. Despotovic 
Outtweeting the Twitterers - predicting information cascades in microblogs.
{\it Microblogs 3rd Workshop on Online Social Networks, WOSN} (2010)
\bibitem{Hu2009osn_official} H.-B. Hu and X.-F. Wang 
{\it Europhysics Letters} 86, 18003 (2009)
\bibitem{MapEquation} M. Rosvall and C. T. Bergstrom
{\it Proc Natl Acad Sci USA} 105, 1118 (2008)
\bibitem{koreanas} H. Y. Yoon and H. W. Park 
Social media information flow and public representation: A case of s. Korean politicians on Twitter
{\it 9th International Triple Helix Conference} (2011)
\bibitem{Barabasi:1999he} A.-L. Barabasi, R. Albert, and H. Jeong 
{\it Physica A: Statistical Mechanics and its Applications} 272, 173 (1999)
\end{thebibliography}
\end{document}